\documentclass[sigconf]{acmart}
\usepackage{amsmath, amsthm}
\usepackage{mdframed}
\usepackage{todonotes}
\usepackage{comment}

\newtheoremstyle{challenge}
  {3pt} 
  {3pt} 
  {\itshape} 
  {} 
  {\bfseries} 
  {.} 
  { } 
  {} 

\theoremstyle{challenge}
\newmdtheoremenv[%
  linecolor=black,
  linewidth=1pt,
  linecolor=black,
  innerleftmargin=10pt,
  innerrightmargin=10pt,
  innertopmargin=6pt,
  innerbottommargin=6pt,
  skipabove=6pt,
  skipbelow=6pt
]{challenge}[theorem]{Challenge}

\AtBeginDocument{%
  }

\copyrightyear{2025}
\acmYear{2025}
\setcopyright{rightsretained}
\acmConference[SIGCSE TS 2025]{Proceedings of the 56th ACM Technical Symposium on Computer Science Education V. 1}{February 26-March 1, 2025}{Pittsburgh, PA, USA}
\acmBooktitle{Proceedings of the 56th ACM Technical Symposium on Computer Science Education V. V. 1 (SIGCSE TS 2025), February 26-March 1, 2025, Pittsburgh, PA, USA}
\acmDOI{10.1145/3641554.3701849}
\acmISBN{979-8-4007-0531-1/25/02}

\begin{document}

\title{Towards s'more connected coding camps}

\author{Ilenia Fronza}
\affiliation{%
  \institution{Free University of Bozen/Bolzano}
  \city{Bolzano}
  \country{Italy}}
\email{ilenia.fronza@unibz.it}
\orcid{0000-0003-0224-2452}

\author{Petri Ihantola}
 \affiliation{%
   \institution{University of Jyväskylä}
   \city{Jyväskylä}
   \country{Finland}}
 \email{petri.j.ihantola@jyu.fi}
 \orcid{0000-0003-1197-7266}

\author{Olli-Pekka Riikola}
\affiliation{%
   \institution{University of Jyväskylä}
   \city{Jyväskylä}
   \country{Finland}}
\email{olli-pekka.v.riikola@jyu.fi} 
\orcid{0009-0009-3958-0595}

\author{Gennaro Iaccarino}
\affiliation{%
  \institution{Direzione Istruzione e Formazione Italiana}
  \city{Bolzano}
  \country{Italy}}
\email{gennaro.iaccarino@scuola.alto-adige.it}
\orcid{0000-0002-7776-7379}

\author{Tommi Mikkonen}
\affiliation{%
  \institution{University of Jyväskylä}
  \city{Jyväskylä}
  \country{Finland}}
\email{tommi.j.mikkonen@jyu.fi}
\orcid{0000-0002-8540-9918}

\author{Linda García Rytman}
\affiliation{%
  \institution{Universitat Jaume I}
  \city{Castellón}
  \country{Spain}}
\email{rytman@uji.es}
\orcid{0009-0003-9785-4052}

\author{Vesa Lappalainen}
\affiliation{%
  \institution{University of Jyväskylä}
  \city{Jyväskylä}
  \country{Finland}}
\email{vesa.t.lappalainen@jyu.fi}
\orcid{0000-0002-3298-0694}

\author{Cristina Rebollo Santamaría}
\affiliation{%
  \institution{Universitat Jaume I}
  \city{Castellón de la Plana}
  \country{Spain}}
\email{rebollo@uji.es}
\orcid{0000-0002-1328-2110}

\author{Inmaculada Remolar Quintana}
\affiliation{%
  \institution{Universitat Jaume I}
  \city{Castellón de la Plana}
  \country{Spain}}
\email{remolar@uji.es}
\orcid{0000-0002-7743-2579}

\author{Veronica Rossano}
 \affiliation{%
   \institution{University of Bari}
   \city{Bari}
   \country{Italy}}
 \email{veronica.rossano@uniba.it}
 \orcid{0000-0002-4079-9641}


\renewcommand{\shortauthors}{Anon et al.}

\newcommand{\oscar}{\texttt{OSCAR}}

\begin{abstract}

%

Coding camps bring together individuals from diverse backgrounds to tackle given challenges within a limited timeframe. Such camps create a rich learning environment for various skills, some of which are directly associated with the camp, and some of which are a result of working as a team during the camp. Unfortunately,  coding camps often remain isolated from the broader educational curriculum or other bigger context, which downplays the opportunities they can offer to students. In this paper, we present the vision of the European initiative \oscar, which aims at connecting coding camps to the educational and professional context faced by the learners. In addition, we sketch a supporting platform and its features for connected coding camps.
\end{abstract}

\ccsdesc[500]{Social and professional topics~Computing education}

\keywords{coding camps, technical platform, requirements analysis, European initiative}


\maketitle

\section{Introduction}
By their nature, coding camps are short-term collaborative activities that bring together individuals from diverse backgrounds to tackle complex challenges within a limited timeframe \cite{porras2019code}. They foster continuous learning \cite{oliveira2022hands}, facilitate networking opportunities \cite{oliveira2022hands}, and cultivate an entrepreneurial mindset \cite{khan2021innovation}. Coding camps provide a rich learning environment that promotes experiential learning by immersing participants in a dynamic, hands-on setting \cite{oliveira2022hands}. Objectives of coding camps often revolve around fostering creative problem-solving, collaboration skills, and, of course, programming~\cite{porras2019code}. These features are also characteristical for hackathons \cite{wilson2019fostering}, and coding camps are sometimes called educational hackathons~\cite{porras2018hackathons}. They broaden participation in computing and engage end-users \cite{decker2015understanding,dewitt2017we}. More recently, coding camps have increasingly been used to teach broader topics such as sustainability~\cite{porras2022experiences} or humanities~\cite{chen2019middle, bryant2019middle}. 
This, together with the fact that coding camps can be executed in various shapes and forms, including online, onsite, and hybrid setups, means that they open new avenues for learning experiences that have long-term impact \cite{wilson2019fostering}.

As in-person collaboration is a key characteristic of coding camps~\cite{pe2018designing}, a wealth of related research provides guidelines on how to organize onsite events \cite{fronza2020end,gama2019developing}. However, the COVID-19 pandemic has changed the traditional setup of coding camps away from onsite. Businesses are now adopting hybrid and remote work models, making it crucial for students to develop skills in this area. Consequently, research has started addressing issues related to online and hybrid delivery formats, such as a lack of a sense of belonging \cite{mooney2021investigating} and fatigue due to prolonged computer use \cite{yousof2021possible}. Yet, a systematic review by Happonen et al.~\cite{happonen2021systematic} found only limited literature on online or hybrid events. Vicinity of participants, face-to-face interactions \cite{pe2018designing}, and fun are key characteristics that help participants advance their technical work, share best practices, and grow individually and collectively in expertise \cite{fronza2022keeping}. Participants reported camps to be more open and inclusive \cite{thayer2017barriers}. However, research in computing education found several barriers participants might face, such as stereotypes of \textit{nerdiness} and \textit{intelligence} exist \cite{thayer2017barriers}, and the need for perseverance and confidence \cite{thayer2017barriers} to face the intensive activities.

One significant issue with coding camps is that they are often isolated from the broader educational curriculum or other societal context. Literature review by Porras et al. \cite{porras2019code} states that although such events are highly praised, the links between them and actual educational activities are weak. This isolation limits the long-term impact and integration of the skills learned during the camps into students' regular academic and personal development. 

The isolation manifests itself in many ways. First, a challenge is that although coding camps are typically very student-centric and encourage students to work on their own ideas \cite{porras2019code}, the organizers still set the stage by deciding when and how the camps are arranged. This limits the options of what participants can do during the camp.

\begin{challenge} \label{challenge1}
The preparation of coding camps is isolated from participants and schools. Coding camps are constrained by organizers who set the stage (e.g., timeframe, theme, and used technologies). 
\end{challenge}
\vspace{2mm}

In addition, coding camps are often separate from schools and other communities where the participants influence and continue acting after the event. This means the participants may miss the natural community support to continue their projects or leverage their learned skills.

\begin{challenge}
The outcomes of the coding camps are too isolated from other activities organized by the school or other communities to which students could belong.
\end{challenge}
\vspace{2mm}

The authors of this position paper are teachers and researchers working in a European initiative \oscar{} that develops and provides coding camps in multiple European countries. Although many of them have a long history in organizing and teaching coding camps, the vision in the project, that is also presented here, is to reorient current coding camps toward participants' perspective, by connecting them to the educational and professional context faced by the learners. 

\section{Timeline of a typical coding camp}
\label{sec:code-camp}

\begin{figure*}[!ht]
\centering
\includegraphics[width=.8\linewidth]{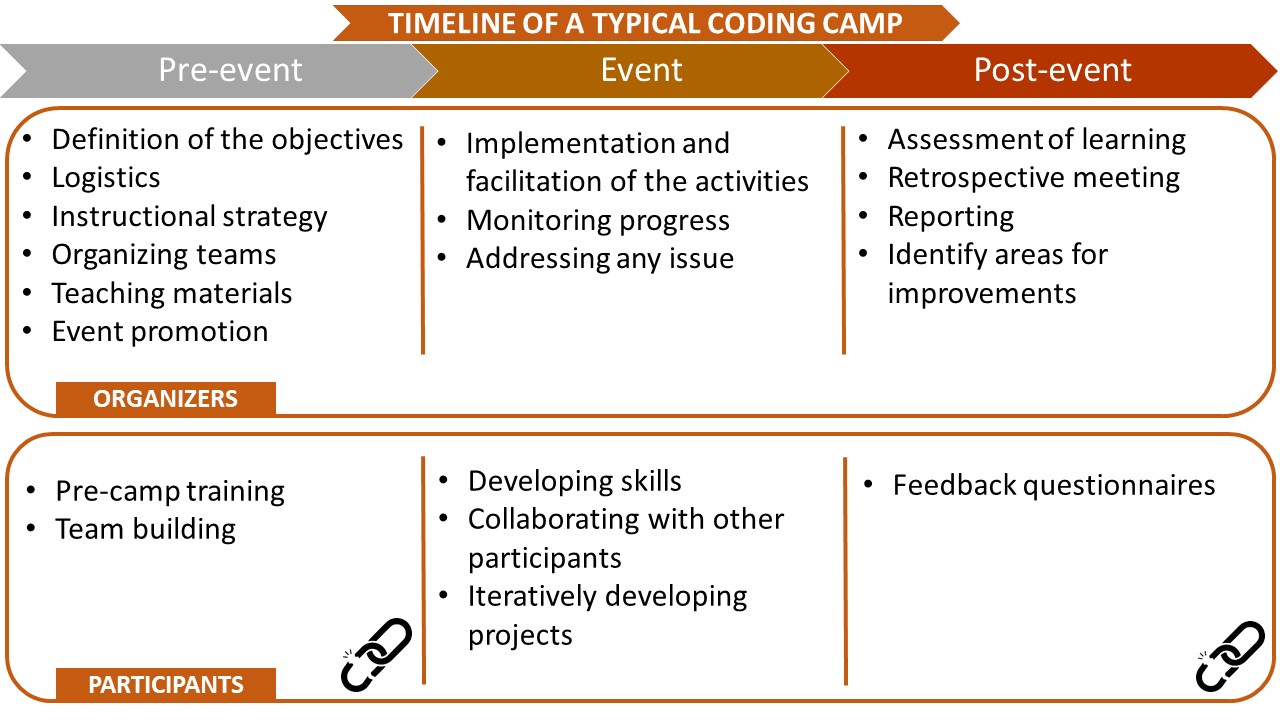}
\caption{Timeline of a \textit{typical} coding camp.}
\label{fig:outline}
\end{figure*}

The challenges with coding camps are related to coding camp timelines, activities, and their relation to stakeholders.
Although a comprehensive taxonomy does not currently exist \cite{porras2019code}, the timeline of coding camps is commonly divided into three main phases: pre-event, event, and post-event \cite{porras2019code}. In the following sections, we outline the typical activities conducted during each phase. As summarized in Figure \ref{fig:outline}, activities can be categorized from both the organizers' and participants' views.

\subsection{Pre-event phase}
During the preparation phase, both organizers and participants set the foundation for a successful coding camp. \textit{Organizers} focus on defining the objectives of the coding camp by specifying, for example, certain programming skills that participants will acquire. Moreover, they handle all the logistics, including arranging the necessary infrastructure such as software, hardware, and communication tools, finding rooms, and defining the schedule. The organizers define all the aspects of the instructional strategy for the coding camp and prepare the necessary materials and resources. Additionally, they promote the event, collect registrations, organize teams, and ensure that the staff involved will be ready for the event. Before the coding camp starts, \textit{participants} can prepare for it by following organizers' instructions, completing any pre-camp training to familiarize themselves with basic concepts and tools used during the camp, and possibly engaging in some team-building activities. However, participants rarely state their learning goals in this phase.

\subsection{Event phase}
This phase is the core of the coding camp, where participants actively engage in learning and project development. \textit{Organizers} facilitate all planned activities, including lectures, hands-on sessions, and collaborative projects. They provide support, troubleshoot, and offer real-time feedback while continuously monitoring progress and adapting the plan as needed. \textit{Participants} acquire new skills and use them in hands-on tasks and projects. They frequently collaborate with others, refining projects in iterations based on feedback.

\subsection{Post-event phase}
The post-event phase is a time of retrospection and assessment. \textit{Organizers} evaluate the participants' learning and distribute certificates of attendance. They also analyze collected data, including participant feedback, to assess the success of the coding camp based on specific metrics. Following a retrospective meeting, they report all the activities, outcomes, challenges, and identify areas for improvement. \textit{Participants} may complete questionnaires to give feedback on the organization, content, and instructional strategy of the coding camp. Typically, communication and support from organizers have been discontinued in this phase. Therefore, participants are not encouraged to reflect on the learning experience, their personal growth, or to identify opportunities for further learning and ways to apply the newly acquired skills. 

As a result, coding camps are constrained by organizers (see Challenge 1) and \textit{not connected} to the rest of the educational activities (see Challenge 2), as shown in Figure \ref{fig:outline} by the broken chain icons in the participants' pre-event, and post-event phases.

\section{Vision: a connected coding camp}
\label{sec:connected-camp} 

The vision of the European initiative involving the authors of this position paper is to enable \textit{connected} coding camps. These camps will support the continuous development of skills from a broader perspective, encourage participants to develop their results further, and become building blocks for planning one’s educational activities. Below, we will define our vision for the learning goals of the coding camps and elaborate on the enhancements and new opportunities made possible by the connected coding camps. Then, we will describe the features of a platform serving as the central tool for implementing the connected coding camps. 

\subsection{Learning goals of a connected coding camp}
\label{subsec:goals}

Coding camps engage participants to identify and solve problems. As such, learning goals can also be designed from a broader perspective. Table \ref{tab:skills} illustrates a set of cross-cutting skills we are currently integrating into our coding camps, based on the reference frameworks created by the European Commission and the Council of Europe to support the conceptualization of the Key competences and their key terminology, i.e., the integrated DigComp 2.2 framework \cite{Vuorikari2022DigComp} (hereafter referred to as DigComp), the European Skills Agenda \cite{Eu2020esa} (hereafter referred to as ESA), and the guide on education for Sustainable Development Goals \cite{unesco2017sdg} (hereafter referred to as SDG). The underlying policy documents were selected because they cover different aspects of society. DigComp provides examples of knowledge, skills and attitudes that help citizens engage confidently, critically and safely with digital technologies. ESA is a five-year plan to help individuals and businesses develop and use more and better skills. SDG identifies indicative learning objectives and suggests topics and learning activities to achieve sustainable development.

\begin{table}[!htb]
  \caption{Relevance to policy contexts of the skills promoted by the coding camps of the European initiative \oscar.}
  \label{tab:skills}
  \begin{tabular}{p{4.5cm}ccc}
    \toprule
    Skills&DigComp&ESA&SDG\\
    \midrule
    Problem solving & x& x& x\\
    Communication and collaboration & x& x& x\\
    Digital content creation & x& & \\
    Entrepreneurial/transversal skills& & x& \\
    Civic/environmental competency&x&x&\\
  \bottomrule
\end{tabular}
\end{table}

Based on the underlying policy documents, the skills in Table \ref{tab:skills} can be defined as follows.

\textit{Problem-solving} includes identifying and solving technical problems when operating devices and using digital environments, assessing needs, and identifying, evaluating, selecting, and using digital tools and possible technological responses to solve them. Furthermore, problem-solving involves creatively using digital tools and technologies to create knowledge and innovate processes and products while engaging individually and collectively in cognitive processing to understand and resolve conceptual problems and problem situations in digital environments.

\textit{Communication and collaboration} include the ability to interact and share data, information, and digital content with others through appropriate digital technologies. Moreover, it involves participating in society using public and private digital services by seeking opportunities for self-empowerment and participatory citizenship through appropriate digital technologies. 

\textit{Digital content creation} involves developing, integrating, and re-elaborating digital content. Moreover, it includes the programming competence, i.e., the ability to plan and develop a sequence of understandable instructions for a computing system to solve a given problem or to perform a specific task.

\textit{Entrepreneurial skills} involve acting upon opportunities and ideas and turning them into (financial, cultural, or social) value for others.

\textit{Civic and environmental skills} include the awareness of the environmental impact of digital technologies and their use and the ability to develop solutions for society with climate considerations. 

Our rationale for the selected approach is that by describing the key competencies and learning goals from a broader perspective, more engaging coding camps can be organized. Moreover, the selected approach addresses the challenges identified in Section 1 by improving the possibility of finding links to other activities.

\subsection{Timeline of a connected coding camp}

\begin{figure*}[!ht]
\centering
\includegraphics[width=.8\linewidth]{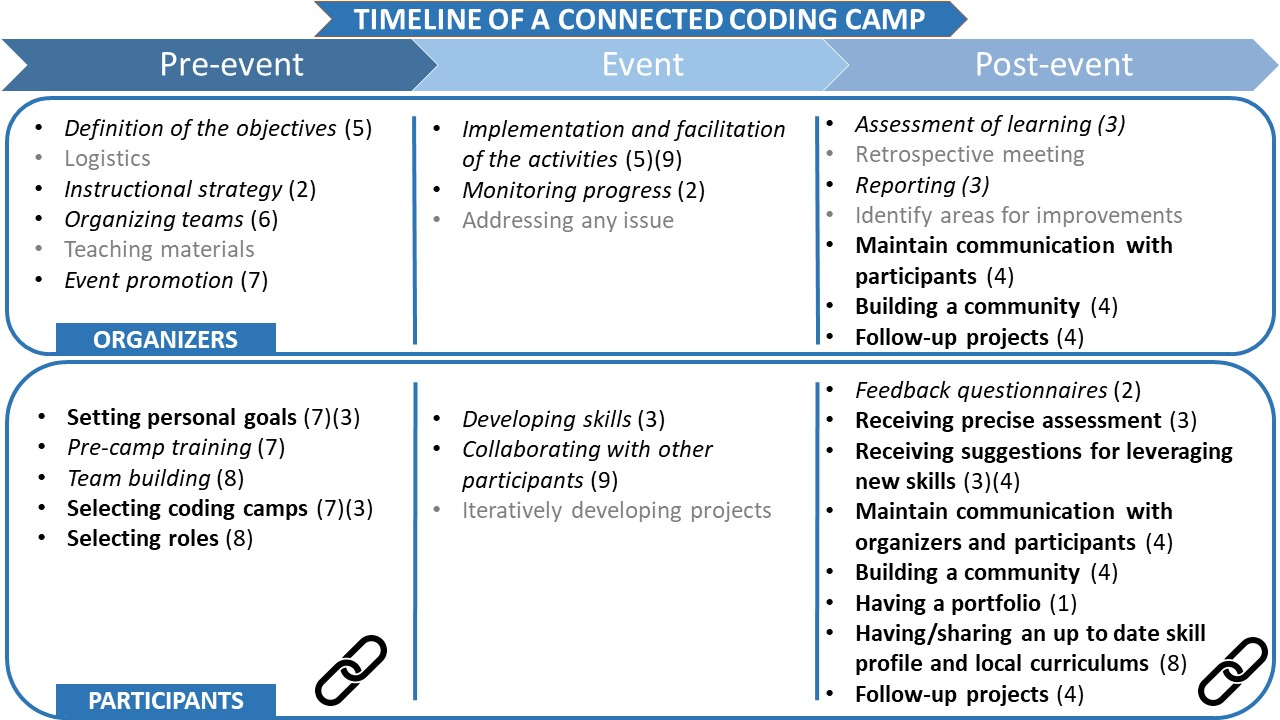}
\caption{Timeline of a \textit{connected} coding camp. Highlighted in italic type are the activities improved compared to a typical coding camp; highlighted in bold are the new possibilities unlocked by the connected coding camp. The numbers in parenthesis after each activity indicate the enabling features of the platform, as detailed in Section \ref{subsec:platform}.}
\label{fig:outline-connected}
\end{figure*}

Figure \ref{fig:outline-connected} illustrates the timeline of a connected coding camp. Below we describe the activities carried out by the organizers and participants in the three phases by highlighting improvements and new possibilities unlocked by connected coding camps in comparison to the typical ones described in Figure \ref{fig:outline}.

\subsubsection{Pre-event phase: Setting goals and expectations} 

In this phase, most of the \textit{organizers}' activities would be improved by our vision. For example, organizing teams would be improved by having data of participants' prior skill levels available. Also, event promotion could be improved by letting participants submit pre-camp training tasks. Organizers could plan coding camps by considering feedback from organizations to better fulfill their needs. Additionally, the objectives of a coding camp could be defined based on the reports of previous coding camps. \textit{Participants} could benefit their coding camp more by  
being active for setting personal goals and expectations by having acquired skills available and useful to select new coding camps. Moreover, the participants' pre-camp training could be improved by completing exercises or by having the possibility to provide ``older'' projects as a confirmation of skills. For improving team building, the participants could create introductory videos of themselves to start knowing the team members. As a new potential activity, participants could be guided through the selection of new parallel/subsequent coding camps and roles (for example, peer tutors) based on their profile. This way, coding camps and goals would initiate from individuals' needs, not from an authority. Additionally, participants could leverage the gained knowledge from  previous coding camps by improving previous projects or taking an idea or an outcome from the materials as a base for a new work.

\subsubsection{Event: Different learning goals}

The vision of a connected coding camp is focused on enhancing the pre-event and post-event phases but also has a positive impact on the event phase. \textit{Organizers} could conveniently monitor the coding camp implementations as all the data would be stored and visualized on a dedicated platform. Furthermore, overall process of merging different data types, such as surveys, questionnaires, and participants' projects could be improved. There would be a notable enhancement on the implementation and facilitation of coding camps also. Experience could be improved upon by facilitating a better delivery of coding camps in an online or hybrid format, allowing more participants from various cities and even countries to take part and collaborate. Also, several coding camps held in various locations could engage participants in their educational journey, thereby enhancing the connection between the coding camps themselves. Finally, less experienced organizers could have a predefined timeline to benefit from the experience of other coding camps' organizers in planning their event. \textit{Participants} could have a better collaboration in their work during the coding camp, even remotely, enhancing their experience by connecting with individuals from other regions or diverse backgrounds. Additionally, participants could receive formative assessments from the organizers through a dedicated platform, improving their chances of developing skills.
 
\subsubsection{Post-event: Connecting coding camp with the overall educational experience.}
When the event is over, \textit{organizers} could provide a more comprehensive assessment by having all the learning outcomes available in a dedicated platform.
Moreover, they could maintain communication with participants
to help building a community for the participants as a long-term effect. To this end, they could suggest potential follow-up projects, or involve participants in future coding camps with a different role, such as a peer tutor \cite{fronza2020evaluating}. Reporting activities would also be easier with all the data available in one place. In a connected coding camp \textit{participants} could receive a precise assessment of skills from the coding camp to bring along for further educational experiences. Moreover, they could get suggestions for leveraging new skills, continuing previous projects, and accessing earlier projects at any time. To this end, participants could keep in touch with their teammates and the organizers by becoming part of an active community. Additionally, participants could keep track of their skills and refer to them when applying for jobs or taking part in other coding camps that build on their skills. Indeed, participants could present their project results as a part of portfolio and export their skills profile to other contexts, such as social media. They could also join follow-up projects proposed by organizers based on the content of previous coding camp implementations. Both autonomously or institutionally driven projects would allow continuous training and could be a link to a school's curriculum. Finally, the platform could help participants to acknowledge how individual projects are related to their study path. 

To summarize, in the proposed vision, coding camps are \textit{connected} to the rest of the educational activities (see Challenge 2) and participants set the stage of their educational experience (see Challenge 1), as shown in Figure \ref{fig:outline-connected} by the chain icons in the pre-event and post-event phases. We trust this in turn increases their impact in participants' individual learning paths. The next section describes the main feature of a platform that would be needed to enable the connected coding camps.

\subsection{Platform to enable connected coding camps}
\label{subsec:platform}
For us, the central tool to enable the connected coding camps is a tailored platform that provides options for using coding camp results and skills achieved as a collection of personal portfolio that support individual goals. Such goals can include various items, such as finding a job that is associated with certain technology, getting involved with others who are interested in the same technology, or simply improve one's skill set for future use. To this end, the platform could provide options to publish the results of a coding camp for others to review, instructions for further development of the outcomes and the platform could also support the participants to link their results for a portfolio of their own. Moreover, the platform could possibly enable a development of new variations by outcomes of the previously organised coding camps and it could link companies, educational institutions, and other stakeholders involved in coding camps. To help achieving the goal of connected camps, the platform needs the following essential features:
\begin{enumerate}
    \item \textit{Publishing the results of a coding camp as tangible, long lasting outcomes, such as personal GitHub repositories or developer portfolios.} Using the platform, participants could release their results on the platform, which may be published as an executable file with a description, image, or video of an outcome. More than that, participants could build their personal portfolio by using various services of their wish and afterwards link the outcomes from the platform to the portfolio.
    \item \textit{Generating a report of coding camps.} The report should be based on all types of data in the platform, such as surveys, projects, and feedback questionnaires. The report could be visible to organizations to provide feedback and suggest topics for  future camps. 
    \item \textit{Providing a clear visualisation of skills and their assessment (e.g., using badges or a gamification approach)}. A dashboard could summarise the assessment at the individual level in each participant's profile. In the organizers' profile, the dashboard could summarise the assessment for each participant as well as for each camp cohort. Skills and assessments should be defined using common frameworks (Section \ref{subsec:goals}) so that different coding camps could contribute to defining each participant's skills. The dashboard could include suggestions to leveraging new skills and provide indicators of a participant's aptitude for specific academic challenges in higher education. 
    \item \textit{Keeping the connection between organizers and participants alive.} The platform should allow participants and organizers to contact each other via chat or by posting new content on a blog even after the coding camp has ended. Using this communication channels, organizers could provide ideas for follow-up projects and participants could choose among different options.
    \item \textit{Providing a timeline for organizing a coding camp, drawing from past coding camps.} The timeline could suggest activities, exercises, and teaching materials.
    \item \textit{Offering convenient tools for collecting participant details}. Collected information may be necessary for official registration documents and to facilitate the coding camp. For example, participants' backgrounds could be useful for defining teams.  
    \item \textit{Providing public pre-camp exercises.} The exercises could assist participants in finding a suitable coding camp and forming teams with feasible skill levels.
    \item \textit{Providing ability to share skills or other achievements} in a user profile for team building and future collaboration with organizations and institutions. The profile could include details, such as a link to the portfolio, a personal introductory video, and the skill profile. Participants should be able to export their profile to other contexts, such as social media.
    \item \textit{Providing tools for participants to support collaboration} both remote and hybrid.
\end{enumerate}

\begin{figure}[!ht]
\centering
\includegraphics[width=\columnwidth]{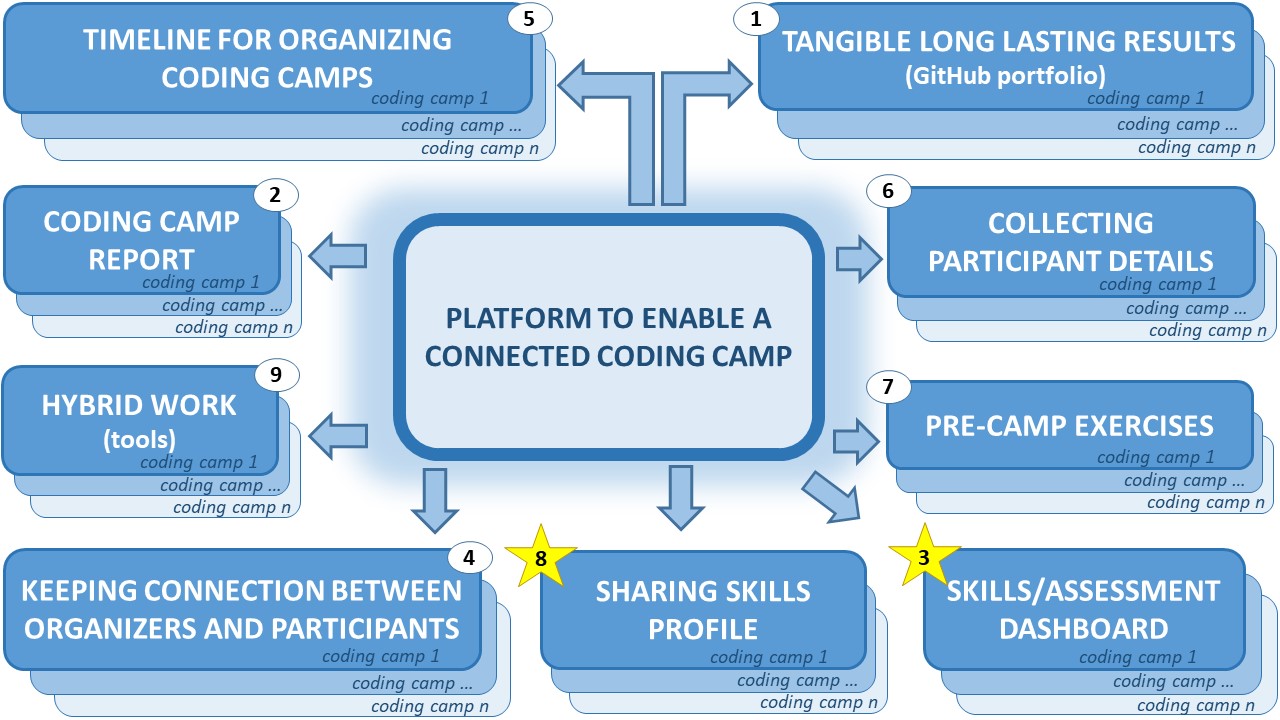}
\caption{Main features of the platform to enable a connected coding camp.}
\label{fig:platform}
\end{figure}

Figure \ref{fig:platform} summarizes the features of the platform. It is important to note that each feature should be based on the input from different coding camps, as represented by the various layers of each feature in the figure. This way, the platform would serve as a strategic hub of connectivity. 

By leveraging its features, the platform enables new activities, which impact the timeline of a \textit{connected} coding camp. Moreover, the fact that several coding camps can share the same infrastructure accumulates data about them to the platform. That way, the students also create a personal portfolio, reflecting their new skills and experiences.

\section{Discussion and Conclusions} \label{sec:discussion}

This position paper's main contribution is to describe coding camps today, discuss how they are isolated from surrounding society, and suggest ways to integrate them with key stakeholders' activities. Although the isolation problem has been identified~\cite{happonen2021systematic}, research on how to tackle it has been scarce. 

One potential reason coding camps remain isolated is that there is not a strong relationship between the activities carried out during the camp and the school's curriculum. Usually, acquired skills during the camp are not included in the student's working portfolio. We argue that it is necessary to establish a link between the coding camp activities, the skills acquired, and the students' curricula and personal interests. Links between the curricula and coding camps are not entirely missing, but in many cases, the participants may just miss seeing them. Thus, connections can be supported both by easing the search for project topics that would be aligned with the curriculum (i.e., pre-event) and by making the hidden connections between the individual projects and curriculum more visible for both participants and schools (i.e., post-event).

Coding camps have multiple interesting technical and organizational challenges, and we call on their organizers, members of the surrounding society (e.g., school teachers), and researchers to collaborate on them. 

The vision of \oscar{} is focused on pre-event and post-event phases (Figure \ref{fig:outline-connected}). Some of the necessary features to implement the vision, discussed in Section, 3 are easier to implement than others. Moreover, some features likely have a more significant impact than others. Because feedback is essential for any learning \cite{hattie2007power}, and because being able to tell others what a participant has done is a prerequisite for many other features, we suggest implementing first the features highlighted with a star in Figure \ref{fig:platform}, i.e., (3) \textit{providing a clear visualization of skills and their assessment} and (8)
\textit{providing the ability to share skills or other achievements}.  Similar features may already be available in modern learning management systems, so if that is the case, we encourage organizers of coding camps to promote their use. 

Although some of the proposed features are relatively straightforward to implement, others may require more effort. Consequently, future research could address the following problems: 

\begin{itemize}

\item \textit{How could AI help participants and schools find connections between the coding camps' topics and the curriculum?} Some development projects already use AI to recognize skills acquired in hobbies (e.g., scouting) as part of formal studies\footnote{see, e.g.,  https://www.osaamiskiekko.fi/en/}, but the research on the topic is limited. In the future, when AI can facilitate learning~\cite{khan2024}, it might be possible to have on-demand coding camps -- events where participants, their teachers, and schools could provide input for a technical platform that would then 1) help participants with similar interests or needs to find each other and 2) help host the event.

\item \textit{What tools are needed and how to improve their usability?} In many cases, ``difficulties in the online environment appear to hinder students to progress'' \cite{happonen2021systematic}. Coding camps differ from many other learning situations by being short-term. More research is needed to better understand how educational software should be integrated into coding camps to get the best out of them (e.g., AI-enabled personal tutors) and simultaneously prevent spending precious time learning new tools.

\item \textit{How to use feedback from coding camps in student guidance or even selection into universities?} The results and level of skills acquired by the student (collected in the platform) can serve as indicators of a student's aptitude for specific academic challenges in higher education. This would complement the information used by schools and teachers to guide students in the decision-making process for selecting a relevant and suitable university program. Moreover, although MOOCs have already been used successfully for university admission \cite{leinonen2019admitting}, the potential role of coding camps as an entry exam has not been explored, and the long-term impact of coding camps has not been studied \cite{fronza2020evaluating}.  
\end{itemize}

Working towards connected coding camps in the framework of a European initiative, the authors of this paper wish to stimulate discussion with respect how to make coding camps first class citizens in the education system. Moreover, we are seeking collaboration with different coding camp stakeholders.

\begin{acks}
This work has been funded by the OSCAR project, number 101132432 ERASMUS-EDU-2023-PI-FORWARD, funded by the European Union. Views and opinions expressed are however those of the author(s) only and do not necessarily reflect those of the European Union. Neither the European Union nor the granting authority can be held responsible for them. 
\end{acks}

\balance
\bibliographystyle{ACM-Reference-Format}
\bibliography{references}


\begin{thebibliography}{25}


\ifx \showCODEN    \undefined \def \showCODEN     #1{\unskip}     \fi
\ifx \showDOI      \undefined \def \showDOI       #1{#1}\fi
\ifx \showISBNx    \undefined \def \showISBNx     #1{\unskip}     \fi
\ifx \showISBNxiii \undefined \def \showISBNxiii  #1{\unskip}     \fi
\ifx \showISSN     \undefined \def \showISSN      #1{\unskip}     \fi
\ifx \showLCCN     \undefined \def \showLCCN      #1{\unskip}     \fi
\ifx \shownote     \undefined \def \shownote      #1{#1}          \fi
\ifx \showarticletitle \undefined \def \showarticletitle #1{#1}   \fi
\ifx \showURL      \undefined \def \showURL       {\relax}        \fi
\providecommand\bibfield[2]{#2}
\providecommand\bibinfo[2]{#2}
\providecommand\natexlab[1]{#1}
\providecommand\showeprint[2][]{arXiv:#2}

\bibitem[Bryant et~al\mbox{.}(2019)]%
        {bryant2019middle}
\bibfield{author}{\bibinfo{person}{Caelin Bryant}, \bibinfo{person}{Yesheng
  Chen}, \bibinfo{person}{Zhen Chen}, \bibinfo{person}{Jonathan Gilmour},
  \bibinfo{person}{Shyamala Gumidyala}, \bibinfo{person}{Beatriz
  Herce-Hagiwara}, \bibinfo{person}{Annabella Koures}, \bibinfo{person}{Seoyeon
  Lee}, \bibinfo{person}{James Msekela}, \bibinfo{person}{Anh~Thu Pham},
  {et~al\mbox{.}}} \bibinfo{year}{2019}\natexlab{}.
\newblock \showarticletitle{A middle-school camp emphasizing data science and
  computing for social good}. In \bibinfo{booktitle}{\emph{Proceedings of the
  50th ACM technical symposium on computer science education}}.
  \bibinfo{pages}{358--364}.
\newblock


\bibitem[Chen et~al\mbox{.}(2019)]%
        {chen2019middle}
\bibfield{author}{\bibinfo{person}{Yesheng Chen}, \bibinfo{person}{Zhen Chen},
  \bibinfo{person}{Shyamala Gumidyala}, \bibinfo{person}{Annabella Koures},
  \bibinfo{person}{Seoyeon Lee}, \bibinfo{person}{James Msekela},
  \bibinfo{person}{Halle Remash}, \bibinfo{person}{Nolan Schoenle},
  \bibinfo{person}{Sarah Dahlby~Albright}, {and} \bibinfo{person}{Samuel~A
  Rebelsky}.} \bibinfo{year}{2019}\natexlab{}.
\newblock \showarticletitle{A middle-school code camp emphasizing digital
  humanities}. In \bibinfo{booktitle}{\emph{Proceedings of the 50th ACM
  Technical Symposium on Computer Science Education}}.
  \bibinfo{pages}{351--357}.
\newblock


\bibitem[Commission et~al\mbox{.}(2020)]%
        {Eu2020esa}
\bibfield{author}{\bibinfo{person}{European Commission},
  \bibinfo{person}{Social~Affairs Directorate-General~for Employment}, {and}
  \bibinfo{person}{Inclusion}.} \bibinfo{year}{2020}\natexlab{}.
\newblock \bibinfo{booktitle}{\emph{European skills agenda – Skills for
  jobs}}.
\newblock \bibinfo{publisher}{Publications Office}.
\newblock
\urldef\tempurl%
\url{https://doi.org/doi/10.2767/978543}
\showDOI{\tempurl}


\bibitem[Decker et~al\mbox{.}(2015)]%
        {decker2015understanding}
\bibfield{author}{\bibinfo{person}{Adrienne Decker}, \bibinfo{person}{Kurt
  Eiselt}, {and} \bibinfo{person}{Kimberly Voll}.}
  \bibinfo{year}{2015}\natexlab{}.
\newblock \showarticletitle{Understanding and improving the culture of
  hackathons: Think global hack local}. In \bibinfo{booktitle}{\emph{2015 IEEE
  Frontiers in Education Conference (FIE)}}. IEEE, \bibinfo{pages}{1--8}.
\newblock
\urldef\tempurl%
\url{https://doi.org/10.1109/FIE.2015.7344211}
\showDOI{\tempurl}


\bibitem[DeWitt et~al\mbox{.}(2017)]%
        {dewitt2017we}
\bibfield{author}{\bibinfo{person}{Anita DeWitt}, \bibinfo{person}{Julia Fay},
  \bibinfo{person}{Madeleine Goldman}, \bibinfo{person}{Eleanor Nicolson},
  \bibinfo{person}{Linda Oyolu}, \bibinfo{person}{Lukas Resch},
  \bibinfo{person}{Jovan~Martinez Salda{\~n}a}, \bibinfo{person}{Soulideth
  Sounalath}, \bibinfo{person}{Tyler Williams}, \bibinfo{person}{Kathryn
  Yetter}, {et~al\mbox{.}}} \bibinfo{year}{2017}\natexlab{}.
\newblock \showarticletitle{What we say vs. what they do: A comparison of
  middle-school coding camps in the cs education literature and mainstream
  coding camps}. In \bibinfo{booktitle}{\emph{Proceedings of the 2017 ACM
  SIGCSE Technical Symposium on Computer Science Education}}.
  \bibinfo{pages}{707--707}.
\newblock
\urldef\tempurl%
\url{https://doi.org/10.1145/3017680.3022434}
\showDOI{\tempurl}


\bibitem[Fronza et~al\mbox{.}(2020a)]%
        {fronza2020end}
\bibfield{author}{\bibinfo{person}{Ilenia Fronza}, \bibinfo{person}{Luis
  Corral}, {and} \bibinfo{person}{Claus Pahl}.}
  \bibinfo{year}{2020}\natexlab{a}.
\newblock \showarticletitle{End-user software development: Effectiveness of a
  software engineering-centric instructional strategy}.
\newblock \bibinfo{journal}{\emph{Journal of Information Technology Education:
  Research}}  \bibinfo{volume}{19} (\bibinfo{year}{2020}),
  \bibinfo{pages}{367--393}.
\newblock
\urldef\tempurl%
\url{https://doi.org/10.28945/4580}
\showDOI{\tempurl}


\bibitem[Fronza et~al\mbox{.}(2020b)]%
        {fronza2020evaluating}
\bibfield{author}{\bibinfo{person}{Ilenia Fronza}, \bibinfo{person}{Luis
  Corral}, \bibinfo{person}{Claus Pahl}, {and} \bibinfo{person}{Gennaro
  Iaccarino}.} \bibinfo{year}{2020}\natexlab{b}.
\newblock \showarticletitle{Evaluating the Effectiveness of a Coding Camp
  through the Analysis of a Follow-up Project}. In
  \bibinfo{booktitle}{\emph{Proc. of the 21st Annual Conf. on Information
  Technology Education}}. \bibinfo{pages}{248--253}.
\newblock


\bibitem[Fronza et~al\mbox{.}(2022)]%
        {fronza2022keeping}
\bibfield{author}{\bibinfo{person}{Ilenia Fronza}, \bibinfo{person}{Luis
  Corral}, \bibinfo{person}{Xiaofeng Wang}, {and} \bibinfo{person}{Claus
  Pahl}.} \bibinfo{year}{2022}\natexlab{}.
\newblock \showarticletitle{Keeping fun alive: an experience report on running
  online coding camps}. In \bibinfo{booktitle}{\emph{Proceedings of the
  ACM/IEEE 44th International Conference on Software Engineering: Software
  Engineering Education and Training}}. \bibinfo{pages}{165--175}.
\newblock


\bibitem[Gama(2019)]%
        {gama2019developing}
\bibfield{author}{\bibinfo{person}{Kiev Gama}.}
  \bibinfo{year}{2019}\natexlab{}.
\newblock \showarticletitle{Developing course projects in a hack day: an
  experience report}. In \bibinfo{booktitle}{\emph{Proceedings of the 2019 ACM
  Conference on Innovation and Technology in Computer Science Education}}.
  \bibinfo{pages}{388--394}.
\newblock
\urldef\tempurl%
\url{https://doi.org/10.1145/3304221.3319777}
\showDOI{\tempurl}


\bibitem[Happonen et~al\mbox{.}(2021)]%
        {happonen2021systematic}
\bibfield{author}{\bibinfo{person}{Ari Happonen}, \bibinfo{person}{Matvei
  Tikka}, {and} \bibinfo{person}{Usman~Ahmad Usmani}.}
  \bibinfo{year}{2021}\natexlab{}.
\newblock \showarticletitle{A systematic review for organizing hackathons and
  code camps in Covid-19 like times: Literature in demand to understand online
  hackathons and event result continuation}. In \bibinfo{booktitle}{\emph{2021
  International Conference on Data and Software Engineering (ICoDSE)}}.
  \bibinfo{pages}{1--6}.
\newblock
\urldef\tempurl%
\url{https://doi.org/10.1109/ICoDSE53690.2021.9648459}
\showDOI{\tempurl}


\bibitem[Hattie and Timperley(2007)]%
        {hattie2007power}
\bibfield{author}{\bibinfo{person}{John Hattie} {and} \bibinfo{person}{Helen
  Timperley}.} \bibinfo{year}{2007}\natexlab{}.
\newblock \showarticletitle{The power of feedback}.
\newblock \bibinfo{journal}{\emph{Review of educational research}}
  \bibinfo{volume}{77}, \bibinfo{number}{1} (\bibinfo{year}{2007}),
  \bibinfo{pages}{81--112}.
\newblock


\bibitem[Khan et~al\mbox{.}(2021)]%
        {khan2021innovation}
\bibfield{author}{\bibinfo{person}{Rakhshanda Khan}, \bibinfo{person}{JUTTA
  HEIKKIL{\"A}}, \bibinfo{person}{Syed Mubaraz}, {and} \bibinfo{person}{Jari
  Luomakoski}.} \bibinfo{year}{2021}\natexlab{}.
\newblock \showarticletitle{Innovation process in business idea generation: a
  case of an entrepreneurial hackathon}.
\newblock \bibinfo{journal}{\emph{EDULEARN21 Proceedings}}
  (\bibinfo{year}{2021}).
\newblock


\bibitem[Khan(2024)]%
        {khan2024}
\bibfield{author}{\bibinfo{person}{Salman Khan}.}
  \bibinfo{year}{2024}\natexlab{}.
\newblock \bibinfo{booktitle}{\emph{Brave New Words: How AI Will Revolutionize
  Education (and Why That's a Good Thing)}}.
\newblock \bibinfo{publisher}{Viking}.
\newblock


\bibitem[Leinonen et~al\mbox{.}(2019)]%
        {leinonen2019admitting}
\bibfield{author}{\bibinfo{person}{Juho Leinonen}, \bibinfo{person}{Petri
  Ihantola}, \bibinfo{person}{Antti Leinonen}, \bibinfo{person}{Henrik Nygren},
  \bibinfo{person}{Jaakko Kurhila}, \bibinfo{person}{Matti Luukkainen}, {and}
  \bibinfo{person}{Arto Hellas}.} \bibinfo{year}{2019}\natexlab{}.
\newblock \showarticletitle{Admitting students through an open online course in
  programming: A multi-year analysis of study success}. In
  \bibinfo{booktitle}{\emph{Proceedings of the 2019 ACM Conference on
  International Computing Education Research}}. \bibinfo{pages}{279--287}.
\newblock


\bibitem[Mooney and Becker(2021)]%
        {mooney2021investigating}
\bibfield{author}{\bibinfo{person}{Catherine Mooney} {and}
  \bibinfo{person}{Brett~A Becker}.} \bibinfo{year}{2021}\natexlab{}.
\newblock \showarticletitle{Investigating the impact of the COVID-19 pandemic
  on computing students' sense of belonging}.
\newblock \bibinfo{journal}{\emph{ACM Inroads}} \bibinfo{volume}{12},
  \bibinfo{number}{2} (\bibinfo{year}{2021}), \bibinfo{pages}{38--45}.
\newblock
\urldef\tempurl%
\url{https://doi.org/10.1145/3408877.3432407}
\showDOI{\tempurl}


\bibitem[Oliveira et~al\mbox{.}(2022)]%
        {oliveira2022hands}
\bibfield{author}{\bibinfo{person}{A. Oliveira}, \bibinfo{person}{H.
  Assumpcao}, \bibinfo{person}{J. Queiroz}, \bibinfo{person}{L. Piardi},
  \bibinfo{person}{J. Parra}, {and} \bibinfo{person}{P. Leitao}.}
  \bibinfo{year}{2022}\natexlab{}.
\newblock \showarticletitle{Hands-on Learning Modules for Upskilling in
  Industry 4.0 Technologies}. In \bibinfo{booktitle}{\emph{IEEE 5th
  International Conference on Industrial Cyber-Physical Systems (ICPS)}}.
  \bibinfo{pages}{1--6}.
\newblock


\bibitem[Pe-Than et~al\mbox{.}(2018)]%
        {pe2018designing}
\bibfield{author}{\bibinfo{person}{Ei~Pa~Pa Pe-Than},
  \bibinfo{person}{Alexander Nolte}, \bibinfo{person}{Anna Filippova},
  \bibinfo{person}{Christian Bird}, \bibinfo{person}{Steve Scallen}, {and}
  \bibinfo{person}{James~D Herbsleb}.} \bibinfo{year}{2018}\natexlab{}.
\newblock \showarticletitle{Designing corporate hackathons with a purpose: the
  future of software development}.
\newblock \bibinfo{journal}{\emph{IEEE Software}} \bibinfo{volume}{36},
  \bibinfo{number}{1} (\bibinfo{year}{2018}), \bibinfo{pages}{15--22}.
\newblock
\urldef\tempurl%
\url{https://doi.org/10.1109/MS.2018.290110547}
\showDOI{\tempurl}


\bibitem[Porras et~al\mbox{.}(2018)]%
        {porras2018hackathons}
\bibfield{author}{\bibinfo{person}{Jari Porras}, \bibinfo{person}{Jayden
  Khakurel}, \bibinfo{person}{Jouni Ikonen}, \bibinfo{person}{Ari Happonen},
  \bibinfo{person}{Antti Knutas}, \bibinfo{person}{Antti Herala}, {and}
  \bibinfo{person}{Olaf Dr{\"o}gehorn}.} \bibinfo{year}{2018}\natexlab{}.
\newblock \showarticletitle{Hackathons in software engineering education:
  lessons learned from a decade of events}. In
  \bibinfo{booktitle}{\emph{Proceedings of the 2nd international workshop on
  software engineering education for Millennials}}. \bibinfo{pages}{40--47}.
\newblock


\bibitem[Porras et~al\mbox{.}(2019)]%
        {porras2019code}
\bibfield{author}{\bibinfo{person}{Jari Porras}, \bibinfo{person}{Antti
  Knutas}, \bibinfo{person}{Jouni Ikonen}, \bibinfo{person}{Ari Happonen},
  \bibinfo{person}{Jayden Khakurel}, {and} \bibinfo{person}{Antti Herala}.}
  \bibinfo{year}{2019}\natexlab{}.
\newblock \bibinfo{title}{Code camps and hackathons in education-literature
  review and lessons learned}.
\newblock , \bibinfo{numpages}{7750--7759}~pages.
\newblock
\urldef\tempurl%
\url{https://doi.org/10.24251/hicss.2019.933}
\showDOI{\tempurl}


\bibitem[Porras et~al\mbox{.}(2022)]%
        {porras2022experiences}
\bibfield{author}{\bibinfo{person}{Jari Porras}, \bibinfo{person}{Eric
  Rondeau}, \bibinfo{person}{Karl Andersson}, \bibinfo{person}{Victoria
  Maria~Palacin Silva}, {and} \bibinfo{person}{Birgit Penzenstadler}.}
  \bibinfo{year}{2022}\natexlab{}.
\newblock \showarticletitle{Experiences from five years of educating
  sustainability to computer science students}.
\newblock In \bibinfo{booktitle}{\emph{Engineering Education for
  Sustainability}}. \bibinfo{publisher}{River Publishers},
  \bibinfo{pages}{1--34}.
\newblock


\bibitem[Riina et~al\mbox{.}(2022)]%
        {Vuorikari2022DigComp}
\bibfield{author}{\bibinfo{person}{Vuorikari Riina}, \bibinfo{person}{Kluzer
  Stefano}, {and} \bibinfo{person}{Punie Yves}.}
  \bibinfo{year}{2022}\natexlab{}.
\newblock \bibinfo{booktitle}{\emph{{DigComp 2.2: The Digital Competence
  Framework for Citizens - With new examples of knowledge, skills and
  attitudes}}}.
\newblock \bibinfo{type}{JRC Research Reports} JRC128415.
  \bibinfo{institution}{Joint Research Centre}.
\newblock
\urldef\tempurl%
\url{https://doi.org/10.2760/115376}
\showDOI{\tempurl}


\bibitem[Thayer and Ko(2017)]%
        {thayer2017barriers}
\bibfield{author}{\bibinfo{person}{Kyle Thayer} {and} \bibinfo{person}{Andrew~J
  Ko}.} \bibinfo{year}{2017}\natexlab{}.
\newblock \showarticletitle{Barriers faced by coding bootcamp students}. In
  \bibinfo{booktitle}{\emph{Proceedings of the 2017 ACM Conference on
  International Computing Education Research}}. \bibinfo{pages}{245--253}.
\newblock
\urldef\tempurl%
\url{https://doi.org/10.1145/3105726.3106176}
\showDOI{\tempurl}


\bibitem[UNESCO(2017)]%
        {unesco2017sdg}
\bibfield{author}{\bibinfo{person}{UNESCO}.} \bibinfo{year}{2017}\natexlab{}.
\newblock \bibinfo{title}{Education for Sustainable Development Goals: learning
  objectives}.
\newblock
\newblock
\urldef\tempurl%
\url{https://unesdoc.unesco.org/ark:/48223/pf0000247444.locale=en}
\showURL{%
Retrieved July 12, 2024 from \tempurl}


\bibitem[Wilson et~al\mbox{.}(2019)]%
        {wilson2019fostering}
\bibfield{author}{\bibinfo{person}{Cait Wilson}, \bibinfo{person}{Thomas
  Akiva}, \bibinfo{person}{Jim Sibthorp}, {and} \bibinfo{person}{Laurie~P
  Browne}.} \bibinfo{year}{2019}\natexlab{}.
\newblock \showarticletitle{Fostering distinct and transferable learning via
  summer camp}.
\newblock \bibinfo{journal}{\emph{Children and Youth Services Review}}
  \bibinfo{volume}{98} (\bibinfo{year}{2019}), \bibinfo{pages}{269--277}.
\newblock


\bibitem[Yousof et~al\mbox{.}(2021)]%
        {yousof2021possible}
\bibfield{author}{\bibinfo{person}{Shimaa~Mohammad Yousof},
  \bibinfo{person}{Rasha~Eid Alsawat}, \bibinfo{person}{Jumana~Ali Almajed},
  \bibinfo{person}{Ameera~Abdulaziz Alkhamesi}, \bibinfo{person}{Renad~Mane
  Alsuhaimi}, \bibinfo{person}{Shrooq~Abdulrhman Alssed}, {and}
  \bibinfo{person}{Iman Mohmad~Wahby Salem}.} \bibinfo{year}{2021}\natexlab{}.
\newblock \showarticletitle{The possible negative effects of prolonged
  technology-based online learning during the COVID-19 pandemic on body
  functions and wellbeing: a review article}.
\newblock \bibinfo{journal}{\emph{Journal of Medical Science}}
  \bibinfo{volume}{90}, \bibinfo{number}{3} (\bibinfo{year}{2021}),
  \bibinfo{pages}{e522--e522}.
\newblock
\urldef\tempurl%
\url{https://doi.org/10.20883/medical.e522}
\showDOI{\tempurl}


\end{thebibliography}


\end{document}